\documentclass[runningheads]{llncs}
\usepackage[T1]{fontenc}
\usepackage{graphicx}
\usepackage{multirow}
\usepackage{color}
\usepackage{comment}
\begin{document}
\title{Are Longer Prompts Always Better? \\
Prompt Selection in Large Language Models for Recommendation Systems}
\titlerunning{On Prompt Selection in LLM-RSs}

\author{Genki Kusano, Kosuke Akimoto, Kunihiro Takeoka}
\authorrunning{Kusano at al.}
\institute{
NEC Corporation \\
\email{\{g-kusano,kosuke\_a,k\_takeoka\}@nec.com}
}

\maketitle              % typeset the header of the contribution
\begin{abstract}
In large language models (LLM)-based recommendation systems (LLM-RSs), accurately predicting user preferences by leveraging the general knowledge of LLMs is possible without requiring extensive training data. By converting recommendation tasks into natural language inputs called prompts, LLM-RSs can efficiently solve issues that have been difficult to address due to data scarcity but are crucial in applications such as cold-start and cross-domain problems. However, when applying this in practice, selecting the prompt that matches tasks and data is essential. Although numerous prompts have been proposed in LLM-RSs and representing the target user in prompts significantly impacts recommendation accuracy, there are still no clear guidelines for selecting specific prompts.

In this paper, we categorize and analyze prompts from previous research to establish practical prompt selection guidelines. Through 450 experiments with 90 prompts and five real-world datasets, we examined the relationship between prompts and dataset characteristics in recommendation accuracy. We found that no single prompt consistently outperforms others; thus, selecting prompts on the basis of dataset characteristics is crucial. Here, we propose a prompt selection method that achieves higher accuracy with minimal validation data. Because increasing the number of prompts to explore raises costs, we also introduce a cost-efficient strategy using high-performance and cost-efficient LLMs, significantly reducing exploration costs while maintaining high prediction accuracy. Our work offers valuable insights into the prompt selection, advancing accurate and efficient LLM-RSs.

\end{abstract}
\section{Introduction}
Recommendation systems are widely used in many modern applications. Their fundamental task is to predict a user's preference for an item. Traditional methods train recommendation systems using supervised learning, leveraging large amounts of user behavioral history as training data \cite{DBLP:conf/recsys/CovingtonAS16,DBLP:conf/recsys/DacremaCJ19,DBLP:journals/computer/KorenBV09}.

In recent years, large language models (LLM)-based recommendation systems (LLM-RSs) have gained significant attention \cite{DBLP:conf/kdd/DeldjooHMKSRVSK24,DBLP:journals/tkde/Fan24_survey,DBLP:conf/coling/LiZLC24,DBLP:journals/tois/Jianghao24_survey,DBLP:journals/www/WuZQWGSQZZLXC24,DBLP:journals/corr/24_survey_prompt_rec}. In LLM-RSs, recommendation tasks are converted into natural language and treated as input to the LLM. This text is called a {\em prompt}, and the LLM generates prediction results on the basis of it. Since LLMs have already acquired a certain level of general knowledge, LLM-RSs can infer a user's preferences for items without training data. Unlike conventional methods, there is no need to gather large amounts of training data or spend long hours training neural networks on expensive machines that utilize GPUs. Therefore, by including task-related text in the prompt, a single LLM can address challenging issues in recommendation systems, such as the cold-start problem \cite{DBLP:conf/nips/VolkovsYP17,DBLP:conf/mm/WeiWLNLLC21,DBLP:conf/www/WangLCCC24} and the cross-domain problem \cite{DBLP:journals/csur/KhanIG17,DBLP:journals/corr/23_cdr_llm,DBLP:conf/wsdm/ZhuTLZXZLH22}. These characteristics lower the barriers to implementing recommendation systems, such as data collection and model development costs, and contribute to research fields and industrial companies with limited data that are eager to adopt recommendation systems.

\begin{figure}[htbp]
\centering
\includegraphics[width=120mm]{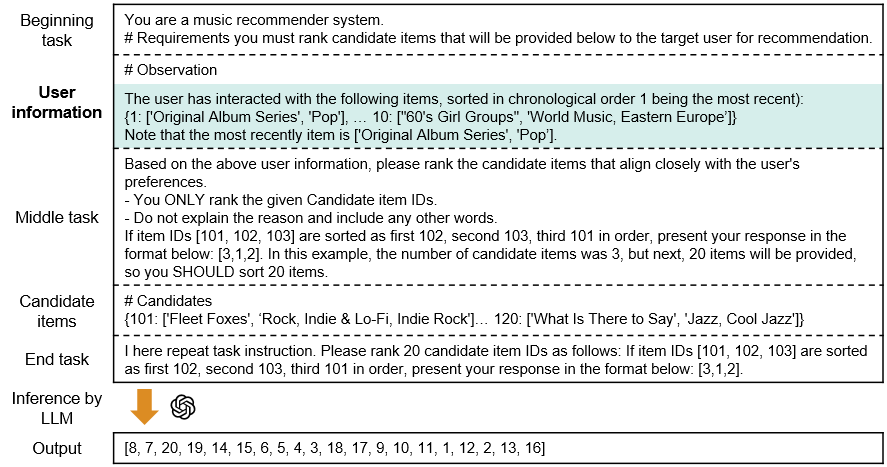}
\caption{An example of a prompt for LLM-RSs. We conduct experiments by varying the user's information part, where the differences from related works are most noticeable.}
\label{fig:prompt-intro}
\end{figure}

A typical prompt for LLM-RSs is: ``The target user has viewed \{item1, ..., item10\} in the past. Please rank the recommended candidates \{item101, ..., item120\} in the order of the user's likely preference'' (Figure \ref{fig:prompt-intro}). The design of prompts significantly influences recommendation results. Although various prompts have been proposed \cite{DBLP:conf/recsys/DaiSZYSXS0X23,DBLP:journals/corr/23_evaluating_gpt,DBLP:conf/cikm/HeXJSLFMKM23,DBLP:conf/ecir/JiLXHGTZ24,DBLP:conf/www/LinSZDCQTY024,DBLP:journals/corr/23_is_chatgpt,DBLP:conf/naacl/PezeshkpourH24,DBLP:conf/recsys/SannerBRWD23,DBLP:journals/corr/23_summary,DBLP:conf/naacl/WangL24,DBLP:conf/naacl/WangJCYZCFLHY24,DBLP:journals/corr/23_llamarec}, their effectiveness varies depending on the specific problem being addressed, making it difficult for users who want to use LLM-RSs to determine the most suitable prompts for their particular scenario. Therefore, this study aims to clarify the guidelines for selecting the most suitable inference prompts for LLM-RSs. 

This paper systematically analyzed previous research prompts to develop an effective method for prompt selection. We found that the significant differences in these prompts, which affect recommendation performance, lie in how user behavioral histories are textualized. We categorized these prompts into four types. Additionally, we focused on the representation of items with which users have interacted in the prompt. While previous research only used item titles in their prompts, we emphasize the importance of categories and descriptions, investigating which combinations contribute most to improving recommendation accuracy.

In the numerical experiments, we conducted 450 trials using 90 types of prompts across five real-world datasets. We observed that no prompts with specific components consistently demonstrated superior performance throughout the experiments. However, we confirmed that adding categories or descriptions improved accuracy under certain conditions. This suggests there is potential to further enhance recommendation accuracy beyond that achieved in previous research utilizing titles.

Despite the challenge of establishing a consistent guideline for prompt selection, we propose a supervised learning method to select suitable prompts for each dataset. Our approach, which is feasible even with a small amount of validation data, selects the prompt with the highest accuracy on the validation data. Compared with four previous studies, our method achieved the highest accuracy in three out of five datasets and the second-highest in one dataset, highlighting the crucial role of prompt selection in recommendation systems. Additionally, we introduce a method that efficiently achieves higher accuracy by leveraging both high-performance and cost-effective LLMs. We can significantly reduce API fees and inference time by using a cost-effective LLM for prompt exploration of the validation data. Our approach achieved the highest accuracy in four out of five datasets and the second-highest in one dataset, demonstrating the potential to reduce exploration costs while maintaining high-accuracy predictions during operation.

Our contributions in this paper are summarized as follows: 
\begin{itemize} 
\item We conducted 450 experiments in LLM-RSs and found that adding categories or descriptions can improve recommendation accuracy depending on the dataset, unlike previous research that used only item titles. 
\item We proposed a method that automatically selects the most suitable prompt for a given dataset and confirmed that it achieves higher accuracy than previous research.
\item We propose a method that reduces exploration costs and improves inference accuracy in the proposed prompt selection method by appropriately utilizing both high-performance and cost-effective LLMs.
\end{itemize}

\section{Related Works}
\label{sec:related_work}
\noindent\textbf{Recommendation and LLM} 
Numerous comprehensive surveys on recommendation systems using LLMs have been conducted in \cite{DBLP:conf/kdd/DeldjooHMKSRVSK24,DBLP:journals/tkde/Fan24_survey,DBLP:conf/coling/LiZLC24,DBLP:journals/tois/Jianghao24_survey,DBLP:journals/www/WuZQWGSQZZLXC24}. LLM-based recommendation systems can be roughly categorized into two approaches: the LLM itself serves as the recommendation system or enhances an existing one. The scope of LLMs varies widely, from models like BERT \cite{DBLP:conf/naacl/DevlinCLT19} and GPT-2 \cite{gpt2_Radford2019LanguageMA} that can be implemented on local machines, to models like ChatGPT and Claude that are accessible only via web APIs. Approaches also diverge on the basis of whether fine-tuning or parameter updates to the LLM's model structure are possible or not. Zhu et al. \cite{DBLP:conf/www/ZhuWGHL24} used GPT-2 and improved accuracy by treating item IDs as unique tokens and modifying the fine-tuning method on the basis of modality (e.g., whether the item has been purchased or just reviewed). Wu et al. \cite{DBLP:conf/www/WuZSYHL24} introduced a method to handle system cold-start situations where there is no user's behavioral history, using the LLM to predict whether a user likes or dislikes an item as a zero-shot classification. The approach in this paper treats the LLM as the recommendation system itself, using only web APIs without fine-tuning. This is motivated by the desire to avoid the need for expensive infrastructure, such as systems utilizing GPUs, in a self-hosted environment.

\noindent\textbf{Prompting for Recommendation} 
Next, we discuss prompts, which are the inputs to LLMs. Several surveys on prompt engineering in general settings have been conducted in  \cite{DBLP:journals/corr/abs-2402-07927_prompt_survey_51,DBLP:journals/corr/abs-2407_12994_prompt_survey_0}. Regarding LLM-RSs, Liu et al. \cite{DBLP:journals/corr/23_is_chatgpt} analyzed the recommendation performance of ChatGPT across five types of recommendation tasks through prompts. Li et al. \cite{DBLP:journals/corr/23_evaluating_gpt} also evaluated ChatGPT, using techniques such as Chain-of-Thought and reranking. Ji et al. \cite{DBLP:conf/ecir/JiLXHGTZ24} tackled sequential recommendation using LLaMA \cite{DBLP:journals/corr/abs-2302-13971_llama}. Wang and Lim \cite{DBLP:journals/corr/23_summary} compressed user information by utilizing summarization prompts for a recommendation. Hou et al. \cite{DBLP:conf/ecir/HouZLLXMZ24_sort} explored ways to improve performance by limiting the items in the user's behavioral history to only the most recent ones. Wang et al. \cite{DBLP:conf/naacl/WangL24} used various techniques, including adding multiple demonstrations from previous inference prompts and their outputs. This approach created prompts with significantly longer token lengths than before. 

\section{LLM-RSs}
\label{sec:llmrec}
In this section, we introduce the task setting, evaluation in LLM-RSs, and the prompts proposed by previous research. To fairly evaluate the effectiveness of each prompt in subsequent experiments, we will standardize the format in Section \ref{subsec:task}, except for the user information part. The differences in the user information found in previous research will be reviewed in Section \ref{subsec:user_info}.

\subsection{Task Description}
\label{subsec:task}
The inference prompts used in this study are fixed to the wording shown in Figure \ref{fig:prompt-intro}, except for the user information part. The basic structure of this prompt is on the basis of that proposed by \cite{DBLP:journals/corr/23_is_chatgpt}. At the beginning part of the task, we added the phrase ``You are a \{genre\} recommender system'' to clarify the role of the LLM \cite{DBLP:conf/recsys/DaiSZYSXS0X23}. For the middle part of the task, we adopted the writing style from \cite{DBLP:conf/naacl/WangL24}, which includes specific task conditions. In the candidate item part, we place items to predict whether the user will like them. Since the prompts can become long, the LLM may forget the instructions in the middle part of the task; therefore, we reiterate the instructions at the end.

The problem setting is as follows. First, we extract the latest two items from the user's behavioral history and treat these as positive examples. Next, we randomly select 18 items that the user has not reviewed as negative examples. These are combined to create 20 candidate items for each user. We use the user's behavioral history to generate the user information part of the inference prompt, excluding the latest two items. The LLM output is a ranking of the candidate items, such as ``[10,8,17,3,...]'' (please see the output in Figure \ref{fig:prompt-intro}). 

For evaluation, we create a list of candidate items where the positions of positive examples are marked as one and negative examples as zero, such as ``[0,0,1,0,...]''. Then, we calculate the ranking metric nDCG@10 between this zero-one list and the predicted ranking scores, which is used to evaluate the performance of the inference prompt. Since the LLM may fail to follow instructions during inference, we repeat the inference up to 10 times until the accuracy can be calculated. If it fails after ten attempts, we calculate the normalized discounted cumulative gain (nDCG) using random ranking scores.

\subsection{Variation of User Information Prompts}
\label{subsec:user_info}
Here, we introduce various textual representations of user information. As shown in Figure \ref{fig:prompt-intro}, using items included in a user behavioral history is common; however, due to the token limit of LLMs, there are methods to sample a subset of these items. Additionally, there is an approach that summarizes the user's profile on the basis of behavioral history and uses this summary in inference prompts. 

\noindent\textbf{Sampling Items Approach}  
Let $I=\{i_1,\ldots,i_n\}$ be the set of items with which a user has interacted in the past. Here, we categorize the following three methods to sample $k$ items from $I$.

\begin{itemize}
    \item \textbf{Latest-$k$} \cite{DBLP:conf/ecir/HouZLLXMZ24_sort} This method sorts the items interacted with by the user in reverse chronological order and uses the latest $k$ items as the sampling result. In the final line, recency-focused prompting is added to emphasize the newest items with its wording shown in Figure \ref{fig:prompt-intro}.
    \item \textbf{Random-$k$} \cite{DBLP:conf/recsys/DaiSZYSXS0X23,DBLP:journals/corr/23_is_chatgpt}\footnote{Although these papers did not explicitly mention using random sampling, they did not specify a concrete item sampling method either, so we referred them here.} This method randomly samples $k$ items from $I$. Here, we sort them in reverse chronological order, similar to the Latest method.
    \item \textbf{Extract-$k$} \cite{DBLP:conf/acl-deelio/LiuSZDCC22,DBLP:conf/naacl/WangL24}\footnote{In \cite{DBLP:conf/acl-deelio/LiuSZDCC22}, extractive prompts are proposed for general use and are not limited to recommendation tasks. Similarly, \cite{DBLP:conf/naacl/WangL24} suggested extractive prompts for demonstrations in the literature of in-context learning, rather than for user behavioral history. Despite these differences, the ideas are similar enough to be referenced here.} This sampling method depends on the candidate items. First, the items are converted to text and then vectorized using sentence embeddings. After vectorization is complete, the $k$ items with the highest similarity are selected as the sampling result. Specifically, let $s_{ij}$ be the cosine similarity between the vector representation of user interaction item $i \in I$ and candidate item $j$. For each item in the user behavioral history, the maximum similarity is defined as $s_i := \max_{j} s_{ij}$. From the similarities $\{s_i\}_{i \in I}$, the top $k$ items are selected. We also sort them in reverse chronological order.
\end{itemize}

\begin{figure}[htbp]
\centering
\includegraphics[width=122mm]{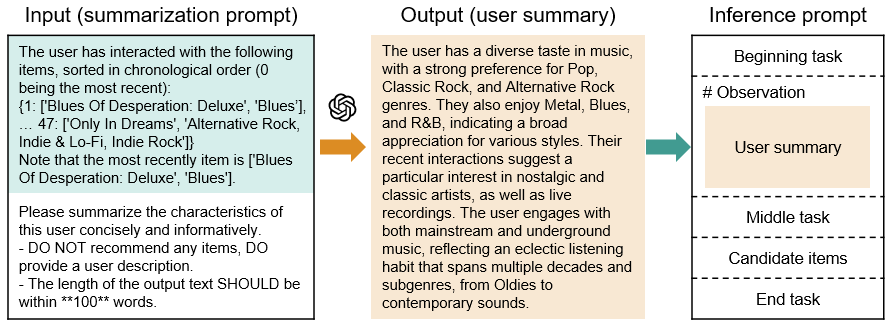}
\caption{Summarization prompt, its output text, and its inference prompt.}
\label{fig:prompt-summary}
\end{figure}

\noindent\textbf{Summarizing Items Approach} 
In previous methods, the collection of items was directly used to represent text in the user information part. In \cite{DBLP:journals/corr/23_summary}, the authors proposed to create a summarized sentence of items for a user and utilize it in LLM-RSs\footnote{The original paper \cite{DBLP:journals/corr/23_summary} performed three LLM inferences: creating a summary, sampling items from the behavioral history that fit the summary, and then recommending items from the candidate list. However, if a user has many items, the token count could exceed the limit during the sampling prompt. Therefore, we modified the process and focused only on the first step.}. As shown in Figure \ref{fig:prompt-summary}, sampled items are included in a summarization prompt to generate a user summary. The generated summary sentence is then placed in the inference prompt.

\section{Experiments}
\subsection{Settings}
\label{subsec:setting}

\noindent\textbf{Dataset}
The data used in the experiments were selected from the Amazon Review Dataset~\cite{DBLP:conf/emnlp/NiLM19}, specifically from the categories of Music, Movies, Books, Grocery, and Clothes. For each dataset, 100 users who had rated 30 or more items with a rating of three stars or higher were selected. The auxiliary information for the items includes three attributes: title, category, and description.

Table \ref{table:data_stat} shows the statistics of the datasets. The title and description columns represent the proportion of items where the token count exceeds or falls below the number in the second row. The category column shows the proportion of items where the number of assigned categories exceeds or falls below the number in the second column. The ``Dup'' column represents the proportion of duplicated items in the sense of being exact matches. The transaction column indicates the number of users who interacted with over 50 or over 100 items.

\begin{table}[htbp]
\caption{Dataset statistics.}
\label{table:data_stat}
\setlength\tabcolsep{3pt}
\centering
\scalebox{1}{
\begin{tabular}{c|ccc|ccc|ccc|ccc}
\hline
 & \multicolumn{3}{c|}{Title} & \multicolumn{3}{c|}{Category} & \multicolumn{3}{c|}{Description} & \multicolumn{2}{c}{Transaction} \\
 & $\leq5$ & $\geq10$ & Dup & $\leq1$ & $\geq3$ & Dup & $=0$ & $\leq5$ & $\geq50$ & $\geq50$ & $\geq100$ \\
\hline
Music & 54.7 & 19.5 & 6.0 & 13.3 & 33.2 & 98.4 & 28.6 & 34.7 & 41.4 & 64 & 39 \\
Movie & 46.6 & 20.9 & 1.2 & 49.2 & 10.3 & 97.9 & 9.6 & 16.0 & \textbf{64.3} & 63 & 28 \\
Grocery & 0.4 & \textbf{93.7} & 0.9 & 1.7 & \textbf{80.9} & 92.7 & 6.7 & 6.8 & \textbf{83.5} & 38 & 12 \\
Clothes & 1.8 & \textbf{74.9} & 4.8 & 0.6 & \textbf{97.4} & 52.1 & 21.3 & 22.4 & \textbf{50.7} & 35 & 8 \\
Book & 25.8 & \textbf{53.9} & 0.8 & 1.5 & 17.2 & 99.1 & 15.1 & 45.5 & 45.7 & 80 & 46 \\
\hline
\end{tabular}
}
\end{table}

These statistics show that the Music and Movie datasets contain less information in their titles and categories. In contrast, the Grocery and Clothes datasets are more informative, with the Book dataset falling somewhere in between.

\noindent\textbf{Models}
We used \texttt{gpt-4o-mini-2024-07-18}\footnote{\url{https://platform.openai.com/docs/models}} as the LLM and set the temperature to 0.3. For sentence embedding in Extract, we used SimCSE \cite{DBLP:conf/emnlp/GaoYC21}\footnote{\url{https://huggingface.co/princeton-nlp/sup-simcse-roberta-large}}.

\noindent\textbf{Prompts}
When creating inference prompts, we can choose from various combinations of titles, categories, and descriptions to represent items. Due to the high proportion of empty descriptions (Table \ref{table:data_stat}), experiments using only descriptions were excluded. By abbreviating the title as T, the category as C, and the description as D, we used the following six combinations: T, C, TC, TD, CD, and TCD. (Figure \ref{fig:prompt-intro} shows the case with TC). For item embeddings in Extract, items are converted into text as follows: T and TD into ``\{title\}'', C and CD into ``\{category\}'', and TC and TCD into ``\{title\}-\{category\}''.

We combined six-item attributes (T, C, TC, TD, CD, TCD) with three formats (Random, Latest, Extract) and four sampling sizes ($k=5,10,20,30$), resulting in 72 prompts. Additionally, summarization prompts for $k=30$ with T, C, and TC and $k=100$ with TD, CD, and TCD across the three formats were included\footnote{When we increased $k$ with TCD, we reached the token limit of \texttt{gpt-4o-mini-2024-07-18} at $k=40$, which is 16,384 tokens.}, resulting in a total of 90 prompts for the experiments. To simplify notation, R-10-T refers to a prompt created using Random-10 and title. Similarly, Latest is denoted as L, and Extract as E. For example, SE-30-TCD refers to the summarization prompt for items sampled by E-30-TCD.

\subsection{Prompts and Accuracy} 

\noindent\textbf{Observation}
For all 90 prompts and five datasets, we calculated the maximum and minimum nDCG@10. From Table \ref{table:score_minmax}, we observed that the prompts achieving the highest accuracy varied in format, sampling size, and item attributes. In terms of accuracy, the minimum can sometimes be as low as the random rate\footnote{When the predicted scores are random, the nDCG@10 empirically distributes around 0.27 in this 2:18 ratio setting, which serves as the minimum baseline.}, indicating the importance of selecting appropriate prompts for each dataset. As seen in the Music, Clothes, and Book datasets, the results also demonstrate the significance of using categories and descriptions, which have not been the focus of previous research.
\begin{table}[htbp]
\setlength\tabcolsep{2pt}
\caption{Maximum and minimum nDCG@10 among the 90 different prompts. ``Acc'' represents accuracy.}
\label{table:score_minmax}
\centering
\scalebox{0.95}{
\begin{tabular}{c|cc|cc|cc|cc|cc}
\hline
& \multicolumn{2}{c|}{Music} & \multicolumn{2}{c|}{Movie} & \multicolumn{2}{c|}{Grocery} & \multicolumn{2}{c|}{Clothes} & \multicolumn{2}{c}{Book} \\
& Prompt & Acc & Prompt & Acc & Prompt & Acc & Prompt & Acc & Prompt & Acc \\
\hline
Max & L-20-TCD & 0.641 & R-10-T & 0.655 & E-5-T & 0.489 & E-5-TCD & 0.563 & L-30-TD & 0.647 \\
Min & SR-100-C & 0.371 & E-5-C & 0.276 & R-10-TD & 0.303 & L-5-C & 0.388 & SE-100-C & 0.379 \\
\hline
\end{tabular}
}
\end{table}

\begin{figure}[htbp]
\centering
\includegraphics[width=120mm]{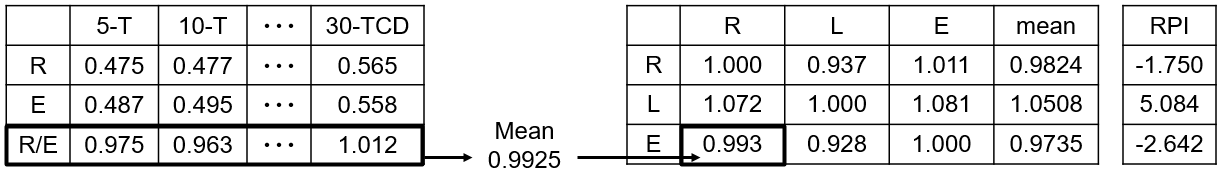}
\caption{Calculation of the relative performance indicator.}
\label{fig:rpi}
\end{figure}

\noindent\textbf{Analysis by Relative Performance Indicator}
From Table \ref{table:score_minmax}, we observed that no specific prompt consistently demonstrated high performance. This raises the question of which components should be used. To evaluate each component quantitatively, we introduce a metric called the relative performance indicator (RPI). As shown in Figure \ref{fig:rpi}, we first fix two components (left) and calculate the accuracy ratio when the other components remain the same. We then average these ratios and create a ratios table across the three formats (right). The RPI is the value obtained by subtracting one from this average and converting it into a percentage. A positive RPI indicates that using the component yields higher accuracy compared with others, suggesting it is a statistically superior component.

Using the sampling items approach, we calculated the RPI for 72 prompts (Table \ref{table:rpi_item}). We observed that either the Extract or Latest format was selected, with a sampling size of 10 or more, and item attributes other than C and CD had high RPIs. Adding more information, such as a large sampling size or including descriptions in TD or TCD, did not always lead to better accuracy. Interestingly, using only categories did not improve accuracy in any dataset, indicating that excluding categories might be more effective. This could be because liking a specific item does not necessarily indicate a preference for other items in the same category. The prompt combinations selected in each RPI analysis showed that, for Music, it matched the maximum accuracy prompt in Table \ref{table:score_minmax}, while for Movie and Clothes, the accuracy was maintained at 90\% of the maximum.
\begin{table*}[htbp]
\caption{RPI of combinations of prompts using the sampling items approach. The ``Prompt'' column indicates the combinations that achieved the best RPI. The ``Ratio'' column indicates the ratio of nDCG@10 of the selected prompt in the ``Prompt'' column to that of the best prompt in Table \ref{table:score_minmax}.}
\label{table:rpi_item}
\setlength\tabcolsep{1.5pt}
\centering
\scalebox{0.95}{
\begin{tabular}{c|ccc|cccc|cccccc|cc}
\hline
& R & L & E & 5 & 10 & 20 & 30 & T & C & TC & TD & CD & TCD & Prompt & Ratio \\
\hline
Music & -1.8 & \textbf{5.1} & -2.6 & -1.9 & -0.3 & \textbf{1.9} & 1.0 & -6.2 & -7.7 & -0.5 & 5.1 & 5.3 & \textbf{6.9} & L-20-TCD & 100 \\
Movie & 1.6 & \textbf{4.7} & -4.6 & -4.0 & -0.1 & \textbf{2.8} & 2.4 & \textbf{24.1} & -26.3 & 3.6 & 13.6 & -7.1 & 12.6 & L-20-T & 95.4 \\
Grocery & -6.4 & 1.9 & \textbf{6.5} & -2.3 & -0.4 & 1.1 & \textbf{3.1} & \textbf{6.7} & -8.2 & 5.0 & 2.5 & -4.3 & 2.1 & E-30-T & 84.7 \\
Clothes & -3.1 & -0.5 & \textbf{4.4} & -1.4 & \textbf{1.9} & 0.9 & -0.7 & \textbf{5.7} & -8.9 & 3.4 & 2.1 & -1.8 & 2.2 & E-10-T & 90.3 \\
Book & -1.2 & \textbf{4.8} & -1.9 & -3.0 & \textbf{2.0} & 0.6 & 1.4 & \textbf{8.8} & -10.9 & 3.5 & 7.5 & -5.7 & 3.3 & L-10-T & 87.1 \\
\hline
\end{tabular}
}
\end{table*}

We also calculated the RPI for summarization prompts (Table \ref{table:rpi_sum}). Unlike in Table \ref{table:rpi_item}, the Extract format was not selected, and the Random format proved more effective in many cases. Given the characteristics of summarization, the Random format may represent a broader range of user interests compared with the Extract format, which selects only specific types of items. Regarding item attributes, we found that including descriptions improved accuracy more than using titles alone.
\begin{table*}[htbp]
\caption{RPI of combinations of prompts using the summarizing items approach.}
\label{table:rpi_sum}
\setlength\tabcolsep{3pt}
\centering
\scalebox{1}{
\begin{tabular}{c|ccc|cccccc|cc}
\hline
 & SR & SL & SE & T & C & TC & TD & CD & TCD & Prompt & Ratio \\
\hline
Music & 3.3 & \textbf{3.7} & -5.9 & -10.5 & -19.2 & -5.5 & 16.1 & 15.0 & \textbf{19.1} & SL-30-TCD & 92.4 \\
Movie & \textbf{3.5} & -0.1 & -2.9 & 5.1 & -26.2 & -0.4 & \textbf{23.1} & -0.6 & 17.5 & SR-30-TD & 92.4 \\
Grocery & \textbf{1.6} & 0.4 & -1.1 & -2.0 & -12.4 & 2.1 & 7.9 & 0.5 & \textbf{8.7} & SR-30-TCD & 88.8 \\
Clothes & 0.4 & \textbf{0.6} & -0.6 & -2.8 & -12.8 & 3.3 & 8.2 & -2.2 & \textbf{11.2} & SL-30-TCD & 97.4 \\
Book & \textbf{1.8} & 1.3 & -2.7 & 3.2 & -20.8 & 3.0 & 12.7 & -3.6 & \textbf{16.2} & SR-30-TCD & 94.0 \\
\hline
\end{tabular}
}
\end{table*}

From Tables \ref{table:rpi_item} and \ref{table:rpi_sum}, it is difficult to identify consistent characteristics that define a better prompt in terms of format or sample size. Regarding item attributes, we observed that using only categories resulted in a low RPI, suggesting that using categories alone should be avoided. Additionally, using only titles was not always the most effective, despite being the focus of previous research. Therefore, we propose that when creating prompts for LLM-RSs, combining titles with categories or descriptions is essential.

\subsection{Exploration of the Prompt with Maximum Accuracy}
\label{subsec:exploration}
We have investigated the characteristics of prompts that contribute to recommendation accuracy. However, before inference, we could not clearly determine which prompt to use for specific conditions or datasets. In a typical supervised learning framework, the prompt that achieves the highest accuracy on validation data should perform well on test data. One of the advantages of LLM-RSs is their ability to perform inference without any training data. In this case, however, we use a small amount of validation data to identify the optimal prompt and then verify whether the selected prompt performs effectively on the test data. For the validation data, we prepared 100 users different from those described in Section \ref{subsec:setting}, using their latest two items as positive examples.

We tested two methods for selecting prompts using validation data. First, we used the method that achieved the highest accuracy on the validation data, referred to as GS, which originates from grid search. Second, we conducted an RPI analysis of the sampling and summarizing items approaches, created two prompts with the best combinations, and selected the prompt that achieved the highest accuracy on the validation data. We also compared the prompts used in previous research as follows: \cite{DBLP:journals/corr/23_is_chatgpt} as R-10-T, \cite{DBLP:journals/corr/23_summary} as SR-100-T, \cite{DBLP:conf/ecir/HouZLLXMZ24_sort} as L-10-T, and \cite{DBLP:conf/naacl/WangL24} as E-30-T.

According to Table \ref{table:score_val}, we confirmed that selecting prompts using validation data, especially those chosen through RPI analysis, achieved high accuracy. Prompts selected by RPI ranked first in three out of five cases and second in one case. For the comparison methods, we observed that while several datasets achieved high accuracy, others experienced a significant drop in accuracy. When comparing GS with prompts selected through RPI analysis, the latter method demonstrated higher accuracy. One reason could be that while GS may yield the highest accuracy on validation data, it often leads to lower accuracy on test data, which can be interpreted as overfitting. In contrast, although the RPI approach might not consistently achieve the highest accuracy on validation data, its components robustly contribute to accuracy, thereby maintaining stable and high accuracy on test data.

\begin{table}[htbp]
\caption{NDCG@10 with selected prompts and those of previous research, where bold and underlined values represent methods ranked first and second in each column.}
\label{table:score_val}
\setlength\tabcolsep{3pt}
\centering
\scalebox{1}{
\begin{tabular}{c|ccccc}
\hline
 & Music & Movie & Grocery & Clothes & Book \\
\hline
R-10-T \cite{DBLP:journals/corr/23_is_chatgpt} & 0.456 & \textbf{0.655} & 0.380 & 0.513 & 0.566 \\
SR-100-T \cite{DBLP:journals/corr/23_summary} & 0.458 & 0.569 & 0.402 & 0.492 & 0.544 \\
E-30-T \cite{DBLP:conf/naacl/WangL24} & 0.499 & 0.547 & 0.414 & \underline{0.535} & \textbf{0.622} \\
L-10-T \cite{DBLP:conf/ecir/HouZLLXMZ24_sort} & 0.527 & \underline{0.646} & \underline{0.417} & 0.523 & 0.564 \\ \hline
\multirow{2}{*}{GS} & \underline{0.590} & \underline{0.646} & \textbf{0.438} & 0.534 & 0.608 \\
& SR-30-TD & L-10-T & E-30-TC & E-30-TD & SR-30-TCD \\
\multirow{2}{*}{RPI} & \textbf{0.624} & 0.624 & \textbf{0.438} & \textbf{0.545} & \underline{0.620} \\
& SR-30-TCD & L-20-T & E-30-TC & SR-30-TD & SL-30-TCD \\
\hline
\end{tabular}
}
\end{table}

\subsection{Inference with High-performance LLM}
\label{subsec:advanced}
We conducted all experiments using \texttt{GPT-4o-mini}. However, we might consider using a more advanced model to achieve higher accuracy. In this section, we investigate how accuracy improves with \texttt{gpt-4o-2024-08-06}, which is more advanced than \texttt{gpt-4o-mini}.

Before using \texttt{gpt-4o} for inference, we review the costs involved in using \texttt{gpt-4o-mini} for the analyses conducted so far. The inferences for 100 test users, with 90 prompts across five datasets, required 84,441 API calls, costing 32.6 USD and taking 29.1 hours. Ideally, it would take $100\times90\times5 = 45000$ API calls, but due to generation failures and up to 10 retries, the processing exceeded 45,000 calls. The preprocessing took an additional 3.9 USD and 4.2 hours for summarization prompts. When applying the same process to validation data, the total cost amounted to 186,415 API calls, 72.3 USD, and 66.7 hours. If the same processes were carried out entirely with \texttt{gpt-4o}, costs would increase more than tenfold, making the use of validation data for prompt search impractical.

To avoid the increased costs of evaluating \texttt{gpt-4o}, we will use only the prompts explored in previous research and the selected prompts listed in Table \ref{table:score_val}. The numerical results are shown in Table \ref{table:score_val_4o}. GS* represents the results using the prompt that achieved the highest accuracy among the four from previous research, GS, and RPI on the validation data with \texttt{gpt-4o}.

From Table \ref{table:score_val_4o}, while the RPI performed well with \texttt{gpt-4o-mini}, it did not perform as well with \texttt{gpt-4o} compared with previous research. Specifically, it underperformed in three out of five datasets compared with L-10-T. However, when recalculating the accuracy of prompts on the validation data using \texttt{gpt-4o} and selecting the most accurate prompt, GS* outperformed previous research in four out of five datasets. Additionally, compared with Table \ref{table:score_val}, there was a notable improvement in overall accuracy. For example, the accuracy in the Book dataset increased significantly from 0.620 to 0.755. Therefore, if the goal is to achieve higher accuracy, using \texttt{gpt-4o} is preferable.

\begin{table}[htbp]
\caption{NDCG@10 with selected prompts and those of previous research, where bold and underlined values represent methods ranked first and second in each column.}
\label{table:score_val_4o}
\setlength\tabcolsep{3pt}
\centering
\scalebox{1}{
\begin{tabular}{c|ccccc}
\hline
 & Music & Movie & Grocery & Clothes & Book \\
\hline
R-10-T \cite{DBLP:journals/corr/23_is_chatgpt} & 0.672 & 0.564 & 0.447 & 0.496 & 0.683 \\
SR-100-T \cite{DBLP:journals/corr/23_summary} & 0.685 & \underline{0.687} & 0.444 & \textbf{0.604} & 0.688 \\
E-30-T \cite{DBLP:conf/naacl/WangL24} & 0.665 & 0.649 & \textbf{0.532} & \underline{0.587} & \underline{0.715} \\
L-10-T \cite{DBLP:conf/ecir/HouZLLXMZ24_sort} & \underline{0.705} & 0.679 & \underline{0.491} & 0.586 & \textbf{0.755} \\ \hline
GS & 0.667 & \textbf{0.706} & 0.475 & 0.577 & 0.682 \\
RPI  & \textbf{0.723} & 0.679 & 0.475 & 0.515 & 0.682 \\
GS* & \textbf{0.723} & \textbf{0.706} & \textbf{0.532} & \underline{0.587} & \textbf{0.755} \\
\hline
\end{tabular}
}
\end{table}

By combining these observations, we can draw practical insights. To maximize accuracy, ideally, 90 prompts should be inferred using \texttt{gpt-4o} on the validation data, and the best-performing one should be selected. However, searching 90 prompts with \texttt{gpt-4o} can be resource-intensive. A more efficient approach is to infer 90 prompts with \texttt{gpt-4o-mini}, select the most effective prompt, and then use that prompt with \texttt{gpt-4o}. This strategy maintains high accuracy during testing while significantly reducing the costs associated with searching prompts on validation data.

\section{Discussion}
\label{sec:discussion}
This study investigated aspects of prompt engineering in LLM-RSs that had not been previously focused on but were found to impact performance significantly. These aspects include item order, sampling size, auxiliary item information, and summarization. In the following sections, we will discuss the potential limitations of our study and areas for future work.

\noindent\textbf{LLM}
In this experiment, we mainly used OpenAI's \texttt{gpt-4o-mini}, and used \texttt{gpt-4o} only in Section \ref{subsec:advanced}. Other LLMs available as API models include Anthropic's Claude and Google's Gemini, while open-source models that can run in local environments include Mistral's Mistral 7B and Meta's LLaMA 2. Since each company is actively competing, their general performance scores and models are updated frequently, and it would require significant time and cost to compare all models fairly. While we anticipate that the tendencies observed in the results evaluated using \texttt{gpt-4o-mini} and \texttt{gpt-4o} would not change significantly with a different LLM, this remains unverified and is a limitation of our study.

\noindent\textbf{Cost in Search Phase}
As discussed in Section \ref{subsec:advanced}, conducting 90 experiments per dataset during the exploration phase with validation data can be costly. While this approach could maximize accuracy, it may be infeasible in budget-constrained scenarios, where excessive spending on exploration can impact inference. Our results suggest that by excluding less effective prompts such as $k=5$, and C, CD, Random for sampling items, and Extract for summarizing items, the number of experiments can be reduced to 36, cutting costs by $40\%=36/90$. However, this reduction may lead to suboptimal prompt selection and potential performance degradation. Thus, future work should focus on finding the best solution to balance exploration and exploitation in LLM-RSs.

\section{Conclusion}
This study examined critical aspects of prompt engineering in LLM-based recommendation systems (LLM-RSs), emphasizing the inclusion of categories and descriptions in prompts, which were previously overlooked. Through 450 experiments across five datasets, we demonstrated that selecting prompts on the basis of dataset characteristics enhances accuracy. We proposed a supervised learning method for automatic prompt selection and introduced a cost-effective strategy with LLMs, significantly reducing exploration costs and improving LLM-RS accuracy.

\newpage

\bibliographystyle{splncs04}
\bibliography{ref}

\end{document}